\documentclass[12pt]{article}
\pdfoutput=1
\usepackage{jheppub}
\usepackage{amsmath}
\usepackage{amsfonts}
\usepackage{amssymb}
\usepackage{graphicx}
\usepackage{fancybox}

\usepackage{amsmath,amssymb,amsthm,cancel}
\usepackage[usenames,dvipsnames]{xcolor}

\usepackage[export]{adjustbox}
\newcommand{\bmat}{\left(\begin{array}}
\newcommand{\emat}{\end{array}\right)}

\def\yzero{\smash{\hbox{$y\kern-4pt\raise1pt\hbox{${}^\circ$}$}}}

\def\beq{\begin{equation}}
\def\eeq{\end{equation}}
\def\beqa{\begin{eqnarray}}
\def\eeqa{\end{eqnarray}}

\def\-{\hphantom{-}}

\def\s2{\frac{1}{\sqrt2}}

\def\beq{\begin{equation}}
\def\eeq{\end{equation}}
\def\beqa{\begin{eqnarray}}
\def\eeqa{\end{eqnarray}}

\def\IF{\relax{\rm I\kern-.18em F}}
\def\II{\relax{\rm I\kern-.18em I}}

\def\Dsl{\,\raise.15ex\hbox{/}\mkern-13.5mu D} 

\def\IS{{\bf {S}}}
\def\IR{{\bf {R}}}
\def\IZ{{\bf {Z}}}

\catcode`\@=11   
\newdimen\@rotdimen
\newbox\@rotbox  

\def\@vspec#1{\special{ps:#1}}
\def\@rotstart#1{\@vspec{gsave currentpoint currentpoint translate
   #1 neg exch neg exch translate}}
\def\@rotfinish{\@vspec{currentpoint grestore moveto}}
\def\@rotr#1{\@rotdimen=\ht#1\advance\@rotdimen by\dp#1%
   \hbox to\@rotdimen{\hskip\ht#1\vbox to\wd#1{\@rotstart{90 rotate}%
   \box#1\vss}\hss}\@rotfinish}
\def\@rotl#1{\@rotdimen=\ht#1\advance\@rotdimen by\dp#1%
   \hbox to\@rotdimen{\vbox to\wd#1{\vskip\wd#1\@rotstart{270 rotate}%
   \box#1\vss}\hss}\@rotfinish}%
\def\@rotu#1{\@rotdimen=\ht#1\advance\@rotdimen by\dp#1%
   \hbox to\wd#1{\hskip\wd#1\vbox to\@rotdimen{\vskip\@rotdimen
   \@rotstart{-1 dup scale}\box#1\vss}\hss}\@rotfinish}%
\def\@rotf#1{\hbox to\wd#1{\hskip\wd#1\@rotstart{-1 1 scale}%
   \box#1\hss}\@rotfinish}%
\def\rotate{\@ifnextchar[{\@rotate}{\@rotate[l]}}
\def\@rotate[#1]#2{\setbox\@rotbox=\hbox{#2}\@nameuse{@rot#1}\@rotbox}

\catcode`\@=12

\begin{document}

\makeatletter
\@addtoreset{equation}{section}
\makeatother
\renewcommand{\theequation}{\thesection.\arabic{equation}}
\pagestyle{empty}
\rightline{IFT-UAM/CSIC-20-169}
\vspace{1.2cm}
\begin{center}
\Large{\bf The Convex Hull Swampland Distance  \\ Conjecture and Bounds on Non-geodesics }
\\[12mm] 

\large{Jos\'e Calder\'on-Infante$^1$,  Angel M. Uranga$^1$, Irene Valenzuela$^{2}$\\[4mm]}
\footnotesize{$^1$Instituto de F\'{\i}sica Te\'orica IFT-UAM/CSIC,
C/ Nicol\'as Cabrera 13-15, 
28049 Madrid, Spain}\\ 
\footnotesize{$^2$Jefferson Physical Laboratory, Harvard University, Cambridge, MA 02138, USA}\\ 
\footnotesize{ 
\href{mailto:j.calderon.infante@csic.es}{j.calderon.infante@csic.es},   
\href{mailto:angel.uranga@csic.es}{angel.uranga@csic.es}, \href{mailto:ivalenzuela@g.harvard.edu }{ivalenzuela@g.harvard.edu }}

\vspace*{5mm}

\small{\bf Abstract} \\
\end{center}
\begin{center}
\begin{minipage}[h]{\textwidth}
The Swampland Distance Conjecture (SDC) restricts the geodesic distances that scalars can traverse in effective field theories as they approach points at infinite distance in moduli space. We propose that, when applied to the subset of light fields in effective theories with scalar potentials, the SDC restricts the amount of non-geodesicity allowed for trajectories along valleys of the potential. This is necessary to ensure consistency of the SDC as a valid swampland criterium at any energy scale across the RG flow. 
We provide a simple description of this effect in moduli space of hyperbolic space type, and products thereof, and obtain critical trajectories which lead to maximum non-geodesicity compatible with the SDC. We recover and generalize these results by expressing the SDC as a new Convex Hull constraint on trajectories, characterizing towers by their scalar charge to mass ratio in analogy to  the Scalar Weak Gravity Conjecture. We show that recent results on the asymptotic scalar potential of flux compatifications near infinity in moduli space precisely realize these critical amounts of non-geodesicity. Our results suggest that string theory flux compactifications lead to the most generic potentials allowing for maximum non-geodesicity of the potential valleys while respecting the SDC along them.
\end{minipage}
\end{center}
\newpage
\setcounter{page}{1}
\pagestyle{plain}
\renewcommand{\thefootnote}{\arabic{footnote}}
\setcounter{footnote}{0}

\tableofcontents

\vspace*{1cm}

\newpage

\section{Introduction}

Since the pioneering work in 
\cite{Vafa:2005ui,Ooguri:2006in,ArkaniHamed:2006dz} and its revival 
\cite{delaFuente:2014aca,Rudelius:2014wla,Montero:2015ofa,Brown:2015iha,Heidenreich:2015wga,Rudelius2015,Heidenreich:2015nta},
 recent years have witnessed an enormous progress in the swampland program, see 
\cite{Brennan:2017rbf,Palti:2019pca} for reviews. The swampland program seeks 
to constrain properties of effective field theories required to be consistently 
realizable in theories of quantum gravity (such as string theory). 
A class of constraints refers to the properties of the moduli space of massless 
scalars in the theory,  in particular the Swampland Distance Conjecture (SDC). A 
loose formulation of its best established variant \cite{Ooguri:2006in} is that, 
as one moves toward infinity in moduli space, there appears an infinite tower 
of states with masses going to zero exponentially with the traversed distance, 
hence below the cutoff of the effective theory, which thus breaks down. The 
strong version of this SDC is that the critical distance controlling this 
exponential scaling is Planckian \cite{Klaewer:2016kiy}. 

The SDC has been extensively tested in string theory compactifications
 \cite{Grimm:2018ohb,Lee:2018urn,Lee:2018spm,Grimm:2018cpv,Corvilain:2018lgw,Joshi:2019nzi,Marchesano:2019ifh,Lee:2019tst,Lee:2019xtm,Lee:2019wij,Baume:2019sry,Blumenhagen_2018,Erkinger:2019umg,Cecotti:2020rjq,Gendler:2020dfp,Klaewer:2020lfg}. Other discussions of this conjecture involve towers of objects other than 
particles 
\cite{Gonzalo:2018guu,Font:2019cxq}, string RG flows \cite{Lanza:2020qmt},
 backgrounds with Anti de Sitter spacetimes 
\cite{Buratti:2018xjt,Lust:2019zwm,Baume:2020dqd,Perlmutter:2020buo}, etc.
The central role of the SDC is also made clear by its connection with other 
conjectures, like the Weak Gravity Conjecture 
\cite{Grimm:2018ohb,Lee:2018spm,Gendler:2020dfp,Lanza:2020qmt,Bastian:2020egp}, the de Sitter conjectures 
\cite{Ooguri:2018wrx,Bedroya:2019snp,Andriot:2020lea,Bedroya:2020rmd}, the $\IZ_k$ Weak Coupling Conjecture 
\cite{Buratti:2020kda}, the emergence proposal 
\cite{Grimm:2018ohb,Heidenreich:2018kpg}, etc.

An important point in the discussion of SDC is that it should apply to {\em 
adiabatic motion} in moduli space. Morally, this amounts to varying the scalar 
values as vevs, i.e. with no spacetime variation. In fact, as shown in 
\cite{Buratti:2018xjt}, backgrounds with spacetime varying scalars can lead to 
transplanckian motion without encountering exponentially falling towers of 
states.\footnote{For spacetime dependence and transplanckian scalar travel, see 
also \cite{Nicolis:2008wh,Draper:2019utz}.} A useful way to understand this point is that 
adiabatic motion corresponds to moving along geodesics in moduli space,  
while 
spacetime dependence introduces extra forces in moduli space motion, leading to 
non-geodesic trajectories. This would seem to imply that the SDC, in the 
adiabatic sense, should apply only to geodesic trajectories in moduli space.

On the other hand, although the SDC and its variants are most precisely stated 
and studied for exactly massless moduli, on physical grounds  they should be 
expected to hold in the presence of scalar potentials, as long as the relevant 
masses and energies remain smaller than the cutoff, i.e. a pseudomoduli space. 
From this perspective, consider a theory with a moduli space ${\cal M}$ 
parametrized by a set of scalars $\phi^i$, such that the SDC is satisfied. If a 
potential $V(\phi)$ is now introduced, the motion of scalars is restricted to 
the valleys of this potential, which can either correspond to left-over 
massless moduli, or to directions along which the potential may not be exactly 
flat but the relevant energies are smaller than a given cutoff $\Lambda$. Let 
us denote this (pseudo)moduli space ${\overline{\cal {M}}}$. At energies below 
$\Lambda$, we may integrate out the heavy directions of ${\cal M}$ and obtain 
and effective theory for the light scalars $\varphi^a$ parametrizing 
${\overline{\cal{M}}}$ .

Now this leads to the following conundrum. In the effective theory below 
$\Lambda$, one can study the SDC by considering geodesic trajectories in the 
moduli space ${\overline{\cal{M}}}$. On the other hand,  the trajectory can be 
regarded as uplifted to a trajectory in ${\cal M}$, so that distances along it 
can be computed as in the parent theory.\footnote{An important point is that 
the kinetic terms on the effective theory in ${\overline{\cal{M}}}$ are 
affected by the integration of the heavy modes; this is captured by the 
statement that in the effective theory below $\Lambda$, the metric on 
${\overline{\cal{M}}}$ is the induced metric from the embedding of 
${\overline{\cal{M}}}\subset {\cal M}$. This is equivalent to the statement 
that the distance is obtained from the trajectory when embedded in ${\cal 
M}$.\label{metric}} But these trajectories in general do not correspond to 
geodesics in ${\cal M}$, and could in principle violate the SDC, {\em even if 
the SCD is obeyed for geodesics in  ${\cal M}$} !

A most relevant aspect of this apparent puzzle is that, if actually realized, 
the SDC would cease to make sense as a swampland constraint. Given an effective 
theory violating the SCD in its moduli space, one could always argue that this 
corresponds to the theory on ${\overline{\cal{M}}}$, and that above certain 
scale $\Lambda$ the theory is completed to a larger moduli space ${\cal M}$ 
which obeys it, and which can in principle be completed into a quantum gravity 
theory. In fact, there is no reason why this cannot occur in a nested manner 
with several effective theories scalating up to some higher energy scale at 
which finally the SDC is fulfilled. In other words, since the notion of moduli 
scape in the presence of multi-scale potentials is a scale-dependent notion, 
the SDC constraint would only apply in a certain energy regime, but then, {\em 
which} energy regime?

We have guided the reader through this argument to make our main point 
manifest. We propose that the above situation cannot occur in a theory of 
quantum gravity, and that the SDC must apply at any energy scale, namely, in 
any of the effective theories valid at any intermediate energy scale. This has 
the following profound implication: since arbitrary scalar potentials in a 
moduli space ${\cal M}$ can easily lead to subspaces ${\overline{\cal{M}}}$ 
violating the SDC, the validity of the SDC at all scales in quantum gravity 
theories constitutes a non-trivial constraint on consistent potentials in 
quantum gravity theories. 

The realization of the SDC in moduli spaces of light fields in the presence of 
potentials has been explored in diverse examples in  flux compactifications in 
string theory e.g. \cite{Baume:2016psm,Valenzuela:2016yny,Blumenhagen:2017cxt}. These top-down approaches are valuable, yet very model dependent. 
In this paper we instead initiate a model-independent bottom-up approach, 
closer to the spirit of the swampland program. Our strategy is instead to 
characterize the non-geodesic trajectories which are nevertheless `sufficiently 
geodesic' to allow the realization of the SDC. The approach is very model 
independent, since it only involves geometrical properties of the moduli space, 
and some information about the towers hiding at its asymptotic regions. 
Characterization of the non-geodesicity allowed by the SDC for trajectories in 
a moduli space, leads in interesting examples to explicit bounds. These bounds 
can be subsequently tested against concrete models, and are interestingly 
saturated in string theory flux compactification in the asymptotic limits \cite{Grimm:2019ixq}. 

The model-independent approach allows us to device an illuminating rephrasing of the SCD in terms of a Convex Hull condition, similar in spirit to that arising in the context of the Weak Gravity Conjecture \cite{Cheung:2014vva}, or its scalar WGC (SWGC) extensions \cite{Palti:2017elp,Gonzalo:2019gjp,Freivogel:2019mtr,DallAgata:2020ino,Benakli:2020pkm,Gonzalo:2020kke}. In particular, we characterize SDC towers by a scalar charge to mass ratio as in the SWGC; this controls the exponential decay rate along asymptotic trajectories characterized in terms of their asymptotic unit tangent vectors. Conversely, by considering the space of such vectors for all possible trajectories, we define an `extremal region' by the set of charge to mass ratios ensuring a fixed minimum decay rate along any possible trajectory. Although in the original formulation of the SDC \cite{Ooguri:2006in}, the decay rate is an undetermined $\mathcal{O}(1)$ factor, concrete lower bounds have been proposed in \cite{Grimm:2018ohb,Gendler:2020dfp,Bedroya:2019snp,Andriot:2020lea,Bedroya:2020rmd,Lanza:2020qmt}. This allows to express the SDC in a given physical system as the condition that the convex hull of the scalar mass to charge ratios of its towers contains the extremal region. If the convex hull condition is not satisfied for arbitrary trajectories, one can use the convex hull to recover the above mentioned bounds on the non-geodesicity of the trajectories. Alternatively, it can also be used to predict the existence of new towers.

The fact that the scalar charge to mass ratio in our Convex Hull SDC agrees with the SWGC is a tantalizing hint, although the physical requirement of `extremality' in both situations does not seem to be necessarily identical. It would be interesting to explore the relationship between the WGC states and the SDC towers in this Convex Hull context, possibly along the lines of \cite{Lee:2018spm,Gendler:2020dfp}.

\medskip

The paper is organized as follows. In Section \ref{sec:hyp} we consider the example of non-geodesic trajectories in a moduli space given by one hyperbolic plane (section \ref{sec:hyperbolic}) or products thereof (section \ref{subsec:two-hype}), and derive bounds on the non-geodesicity of trajectories obeying the SDC.  The analysis of the multi-moduli cases motivates Section \ref{sec:subspaces}, where we frame the multi-axion examples in a general reformulation of the SDC (section \ref{sec:subspaces}). This allows us in Section \ref{sec:convex-hull} to formulate our Convex Hull SDC (section \ref{sec:formulation-sdc}), and recover and vastly generalize results of the previous sections, as we show in several explicit examples (section \ref{sec:chsdc-examples}). In Section \ref{sec:asymptotic-flux} we revisit the results about asymptotic flux compactifications in \cite{Grimm:2019ixq}, and show that they realize the critical behaviours of non-geodesicity. Section \ref{sec:conclu} contains our final remarks.

\section{Non-geodesic bounds in the hyperbolic plane}
\label{sec:hyp}

We focus our analysis on trajectories approaching points at infinity in moduli 
space, in the spirit of the SDC, since the interesting physics occurs in the 
asymptotic region near infinity. Also, it often corresponds to weakly coupled 
regimes, where effective actions and scalar potentials can be reliably 
computed. Moreover, fairly general moduli spaces simplify in the asymptotic 
regime, so that very simple moduli space geometries are useful templates for the asymptotics 
of general moduli spaces.

In this section, we discuss moduli spaces given by a hyperbolic plane, or products thereof. Despite their apparent simplicity, they are key to describing moduli spaces of general CY compactifications near their boundaries at infinity \cite{Grimm:2018ohb,Grimm:2018cpv,Grimm:2019ixq}, to the extent of encoding much of the dynamics of these models \cite{Grimm:2020cda}. Moreover, they allow for explicit computations which will be useful to motivate our generalizations in later sections.

\subsection{One hyperbolic plane}
\label{sec:hyperbolic}

Consider a 4d effective theory with two real moduli $s$ and $\phi$ with kinetic terms
\beqa
\frac{n^2}{s^2}\, (\, \partial_\mu\, s\, \partial^\mu \,s\, +\, \partial_\mu\, \phi\, \partial^\mu \,\phi\,) \, ,
\eeqa
where $n$ is a free parameter. In other words, the moduli space ${\cal M}$ is 
given by the upper half-plane with metric 
\begin{equation}
    d\Delta^{2}=\frac{n^{2}}{s^{2}}\left(ds^{2}+d\phi^{2}\right). 
    \label{hyperbolic-plane}
\end{equation} 
We note that $n$ determines the Ricci scalar curvature
\begin{equation} 
    R=-\frac{2}{n^{2}}.
\end{equation}
This geometry is ubiquitous in string theory, with $\phi$ corresponding to some 
periodic axion and $s$ its `saxion' partner (although we do not assume susy, we 
stick to this name). For instance, the type IIB complex coupling in 10d, the 4d 
axio-dilaton in string compactifications, and the K\"ahler and complex 
structure moduli of 2-tori in toroidal (and orbifold and orientifolds thereof) 
compactifications. 

In many of these, the $SL(2,\IR)$ symmetry of the above geometry lead to an 
exact infinity discrete $SL(2,\IZ)$ duality symmetry. However, we work in a 
more general perspective, so that our analysis is valid in the absence of this 
symmetry. On one hand, in many compactifications, we would like to regard the above metric as a good approximation 
to the moduli space (or suitable subspaces thereof) of CY compactifications, in the large $s$ asymptotic region; hence the region near $s=0$ is not relevant to this physics context, and  the duality $s \to 0$ and $s \to \infty$ is a mere artifact. Second, the discrete axion periodicity (which would be present even near $s\to\infty$) is in general 
spontaneously
broken in the presence of potentials of axion monodromy\footnote{More precisely, the discrete 
periodicity is preserved due to the multiple-branched structure of the potential. However, it 
is spontaneously broken when the problem under study (e.g. adiabatic 
motion in moduli space) is restricted to a single branch.} kind 
\cite{Silverstein:2008sg,Kaloper:2008fb}, a generic situation in flux 
compactifications \cite{Marchesano:2014mla,McAllister:2014mpa}. Hence we 
consider $\phi$ to take real values, with no identification whatsoever.

We consider that the SDC is satisfied on this moduli space ${\cal M}$, namely 
there exists a tower of states with mass scale\footnote{The fact that the overall scale is independent of $\phi$ does not imply that the masses of individual states in the tower can not depend on $\phi$. Indeed, a typical structure is given by $M_n=M|n+\phi|$, with $n$ labeling states in the tower and $M$ depending only on $s$, as required for our analysis. Here the $\phi$-dependence is determined by the fact \cite{Grimm:2018ohb,Corvilain:2018lgw} that the different states in the tower are generated by monodromy in $\phi$, i.e. $\phi\to\phi+1$ is equivalent to $n\to n+1$. More general axion dependences in the tower scale will be easily included in the analysis in Section \ref{sec:convex-hull}.}
\begin{equation} 
    M \sim s^{-a},\,a>0.
    \label{hyper-tower}
\end{equation}

If $s$ parametrizes the vertical axis, all geodesics in this space are either
vertical lines or half-circles with centers on the $s=0$ line. Thus, the only
geodesics approaching $s\to \infty$ are vertical lines with $\phi =
\text{const.}$
For these geodesics, the distance behaves as
\begin{equation} 
    \Delta\,\sim\,n\,\log s
\end{equation} 
(with $n$ taken positive herefrom). The mass scale of the tower reads
\begin{equation}
    M \sim\exp\left(-\frac{a}{n}\Delta\right) \sim 
    \exp\left(-\alpha\Delta\right),
\end{equation} 
thus leading to the SDC with decay rate $\alpha=\frac{a}{n}$. This is indeed $\mathcal{O}\left(1\right)$ in many
realizations in string theory.

Let us now consider a general trajectory approaching $s\to \infty$  in this
moduli space. For reasonable trajectories, we can  can use $s$ to parametrize
it,\footnote{Actually, requiring that the trajectory eventually goes to
$s\to\infty$ makes this parametrization always valid for sufficiently large
$s$.} so that the curve is defined by the expression
\beqa
\phi =f(s) 
\label{single-path}
\eeqa
for some function $f$ that we assume sufficiently smooth. This is  a template 
to describe the moduli space of an effective theory in which there is partial 
moduli stabilization, and the light direction can be parametrized by $s$.

Recalling footnote \ref{metric}, note that the distance in this effective theory
is not measured by just the metric component $g_{ss}$, but rather by the
effective metric obtained upon replacing the $s$-dependent value of $\phi$ in
the underlying metric. This is equivalent to measuring distance along the
trajectory with the ambient space metric (\ref{hyperbolic-plane}) in ${\cal M}$.
This yields
\begin{equation} \label{eq:curve}
   \,d\Delta=\frac{n}{s}\sqrt{1+f^{\prime}
    \left(s\right)^{2}}\,ds.
\end{equation} 

We can now classify general trajectories in three different kinds, according to 
the asymptotic behaviour of $f^{\prime}(s)$ in the $s\to \infty$ limit:
\begin{itemize} 

\item The {\em Asymptotically Geodesic} case:    
    
This corresponds to $f^{\prime}\left(s\right)\rightarrow 0$, and we have
\beqa    
d\Delta\,=\, n \frac{ds}{s} \, .
\eeqa
Trajectories of this class approach a geodesic when $s\to \infty$. Therefore the SDC is automatically satisfied with decay rate 
\beqa
\alpha_{\rm geod.} =\frac{a}{n} \, .
\label{alfa-geod}
\eeqa

\smallskip
   
\item  The {\em Critical} case: 

This corresponds to $f^{\prime}\left(s\right)\rightarrow\beta\,=\,{\rm const.}$ 
and we have
\beqa
d\Delta\,=\,\sqrt{1+\beta^{2}}\,n\,\frac{ds}{s}\, .
\eeqa
The tower of states has mass scale 
\begin{equation}
M \sim\exp\left(-\frac{\alpha}{\sqrt{1+\beta^{2}}}\Delta\right),
\end{equation} 
which is consistent with the SDC, but modifies the scale of the exponential. We can define a factor
\begin{equation} 
\label{eq:nu-factor}
\nu \equiv \frac{\alpha_{\text{geod}}}{\alpha_{\text{non-geod}}}
\end{equation}
that measures such a modification, so that $M\sim \exp(-\frac{\alpha_{\rm geod}}{\nu} \Delta)$. In this case we have 
\beqa
\nu_{\rm crit.}=\sqrt{1+\beta^{2}} \; \rightarrow\; \alpha_{\rm crit}\, = \frac{a}{n\sqrt{1+\beta^2}} \, .
\label{critical-values}
\eeqa
 Hence, $\nu\to\infty$ corresponds to a violation of the SDC in a non-geodesic trajectory.

\smallskip
    
\item  The {\em Swampy} case:   

This corresponds to $f^{\prime}\left(s\right)\rightarrow\infty$, and we have
\beqa
d\Delta\,=\,   n 
\frac{\left|f^{\prime}\left(s\right)\right|}{s}ds \, .
\eeqa
Here we can evaluate the behaviour of the tower by computing 
 \begin{equation}
\frac{d\log M}{d\Delta}=\frac{d\log M}{ds}\frac{ds}{d\Delta} =
        -\frac{a}{n}\left|f^{\prime}\left(s\right)  \right|^{-1}\rightarrow0.
    \end{equation} 
This violates the SDC since the tower mass scale is no longer falling 
exponentially with the distance.
    
\end{itemize} 

We note that the critical case corresponds to the maximum deviation from a 
geodesic still consistent with the SDC. It therefore provides a non-trivial 
bound that any light direction in the moduli space (after partial moduli 
stabilization by some scalar potential) must obey.
Notice that this is the class of curves in which the saxion varies linearly 
with the axion ($s \sim a$). This will be contrasted with explicit string 
models of flux compactifications in Section \ref{sec:asymptotic-flux}, where we 
show that this class of models saturates the bound.

\medskip

It is interesting to explore the characterization of these classes in terms of 
a geometrical quantity of the trajectories. A natural scalar measure of 
non-geodesicity is the modulus $|\Omega|$ of the proper 
acceleration\footnote{Actually, since our trajectories are not worldlines, we should use the term ``extrinsic curvature'', but we stick to the kinematical language.}
\beqa \label{proper-acceleration}
\Omega^{i}\, =\, T^{j} \nabla_{j} T^{i} \, ,
\eeqa
where $T$ is the normalized tangent vector of the trajectory and $\nabla$ is de 
covariant derivative in moduli space. In our case we have
\begin{equation}
    |\Omega|^{2} = \frac{\left( f^{\prime}(s) + f^{\prime}(s)^{3} - sf^{\prime 
    \prime }(s) \right)^{2}}{n^{2} \left( 1 + f^{\prime }(s)^{2}\right)^{3}}.
\end{equation}

We can now characterize the three classes of paths above in terms of the asymptotic behaviour for $|\Omega|$ as follows.  The  {\em Asymptotically Geodesic} case corresponds to $f^{\prime} \left( s\right) \rightarrow 0 $ which translates to $ |\Omega|^{2} \rightarrow 0 \, $. The proper acceleration vanishes asymptotically, since the trajectory approaches a geodesic. Contrary, for the  {\em Swampy} case, one has
$f^{\prime} \left(s\right) \rightarrow \infty $ leading\footnote{One also needs that $\frac{sf^{\prime\prime}}{f^{\prime 3}}\to 0$. This is satisfied for all functions of the form $f(s)=s^{n},(\log s)^{n},e^{s^{n}}$.} to $ |\Omega|^{2} 
\rightarrow \frac{1}{n^{2}} $.
Thus, the proper acceleration at $s \rightarrow \infty $ attains a maximal value,  
and signals a hard violation to the SDC. Finally, the {\em Critical} case corresponds to
\beqa
f^{\prime} \left( s\right) \rightarrow \beta={\rm const.} \, \Longrightarrow \, 
|\Omega|^{2} \rightarrow \frac{1}{n^{2}} \frac{\beta^{2}}{1 + \beta^{2}} \, .
\eeqa
The change in the parameter of the exponential defined in \eqref{eq:nu-factor} can be written as    
\begin{equation} 
\label{eq:RSDC-violation} 
\nu = \frac{1}{\sqrt{1-n^{2}|\Omega|^{2}}}. 
\end{equation}
We can thus recast the criterion that the trajectory respects the SDC as a 
bound on the non-geodesicity:
\begin{equation} 
\label{eq:criterium} 
     |\Omega|^{2} \, <\,
    \frac{|R|}{2}=\frac{1}{n^{2}}
\end{equation}

In the next subsection, we will generalize these bounds to higher dimensional moduli spaces in which there is more than one hyperbolic plane. We will see that they cannot be stated simply in terms of the modulus of the proper acceleration, as the direction will also matter. In other words, the bounds will vary depending on the type of trajectory/ the growth sector to which the trajectory belongs.

\subsection{Product of hyperbolic planes}
\label{subsec:two-hype}

We now consider a moduli space given by a product of hyperbolic planes. For simplicity, we consider the case of two, which suffices to illustrate the point. The metric is
\begin{equation}
    d\Delta^{2}=\frac{n^{2}}{s^{2}}\left(ds^{2}+d\phi^{2}\right) + 
    \frac{m^{2}}{u^{2}}\left(du^{2}+d\psi^{2}\right) \, .
    \label{two-hype-metric}
\end{equation}
As in Section \ref{sec:hyperbolic}, we focus in paths approaching the  infinite distance regime  $s,u \to \infty$. Note that this setup (\ref{two-hype-metric}) nicely models the asymptotic behaviour of CY moduli spaces, cf. Section \ref{sec:asymptotic-flux}, thus our discussion is of direct relevance to the study of non-geodesic paths in CY moduli space.

In order to satisfy the SDC for geodesics, it is enough to introduce a tower of states for each of the hyperbolic planes, with mass scales
\begin{equation}
    M_{s} \sim s^{-a}, \qquad M_{u} \sim u^{-b}, \qquad a,b>0.
    \label{two-towers}
\end{equation}
It is clear that these towers enforce the SDC for geodesic paths contained in a single hyperbolic plane. Moreover, one can show that they also suffice to recover the SDC for more general geodesics.\footnote{Physical realizations like CY compactification may produce additional towers. This is ignored in this section for simplicity, but is included in the general analysis in later sections.}

Let us consider non-geodesic paths approaching infinity in different ways, some of which correspond to different growth sectors, in the terminology of \cite{Grimm:2018cpv}. For instance, we can consider paths that only move on one of the hyperbolic planes, namely approaching $s\to\infty$ while keeping $u,\phi$ fixed, or alternatively, approaching $u\to\infty$ with $s,\phi$ fixed. In that case, the situation is equivalent to the one discussed in section \ref{sec:hyperbolic} and we can simply borrow the classification of asymptotically geodesic, critical and swampy  trajectories. We could also apply the criterion  (\ref{eq:criterium}) on the proper acceleration $|\Omega|^2$, although this would yield two {\em different} bounds for the two different curvature paratemeters $n$ and $m$. In other words, there is no single discriminating criterion on $|\Omega|$ which applies to both kinds of paths.

On the other hand, we may consider a path  
\begin{equation}
	\psi = f(s)\, \quad {\rm with}\; \phi,u\; {\rm const}.
\end{equation}
so that we move along a trajectory involving the axion and saxion of different hyperbolic planes.
The distance along this curve is given by
\begin{equation}
	d\Delta = \sqrt{\frac{n^2}{s^2} + \frac{m^2}{u^2} f^{\prime}(s)^{2}}\, ds = \frac{n}{s} \sqrt{1 + \frac{m^2}{n^2} \frac{s^2}{u^2} f^{\prime}(s)^{2} } \, ds \, .
\end{equation}
In the second equality we recognize the two terms in the square root to be the
geodesic and the non-geodesic contribution to the field distance.

As done in section \ref{sec:hyperbolic}, we can classify the trajectories in the same
three different families, depending on the relevance of the non-geodesic
contribution in the asymptotic limit $s\to\infty$.
We thus see that the critical case is:
\begin{equation}
	\frac{m^2}{n^2} \frac{s^2}{u^2} f^{\prime}(s)^{2} \to \text{const.}\, ,
\end{equation}
which means that $s f^{\prime} \to \gamma = \text{const.}$, implying the critical behaviour
\begin{equation}
	f(s) \to \gamma \log s \, .
\end{equation}
If $f(s)$ grows slower or faster than this critical case one gets the
asymptotically geodesic and the swampy case respectively.

The SDC will only be satisfied for asymptotically geodesic or critical trajectories. Only for those paths, the towers of states in \eqref{two-towers} will still exhibit the exponential behaviour in terms of the distance along the path. Recall that the critical case is the maximum deviation from a geodesic still consistent with the exponential behaviour required by the SDC, although the factor in the exponential changes.
For a tower with $M_s \sim s^{-a}$ one finds 
\begin{equation} \label{eq:non-geodesic-exp-rate}
	M_s \sim \exp \left( - \frac{\alpha}{\nu} \right) \, ,
\end{equation}
with $\alpha=\frac{a}{n}$ being the geodesic decay rate, and the
factor of violation of the SDC given by
\begin{equation}
	\nu = \sqrt{1 + \frac{m^2}{n^2} \frac{\gamma^2}{u^2}} \, .
\end{equation}
In comparison with the case of dual axion-saxion pair, we find that this factor
varies when choosing different $u=\text{const.}$ planes. 

This also produces a direct relation, albeit a different one, between $\nu$ and the modulus of the proper acceleration. Indeed we find
\begin{equation} \label{eq:RSDC-violation2} 
    \nu = \frac{1}{\sqrt{1-m|\Omega|}}. 
\end{equation}

Hence the existence of two axionic directions on which the path can wind lead to different relations between $\nu$ and $|\Omega|$. An implication is that, taking the $\nu \to\infty$ limit, they lead to different discriminating criteria. The critical values correspond to $|\Omega|\to \frac{1}{n}$ and $|\Omega|\to \frac{1}{m}$, when moving along $\phi$ or $\psi$ respectively. 

The difficulty in finding a criterion based solely on the modulus of the proper acceleration is that $|\Omega|$ misses the information about the direction in moduli space. And trying to include this information more explicitly may quickly run into nasty and unphysical dependences on the coordinates chosen in moduli space. In the next section we provide a concrete description which avoids these pitfalls, and yet allows to provide a purely geometrical reformulation of the SDC in general moduli spaces. We will then translate this general criterion into a Convex Hull SDC condition in section \ref{sec:convex-hull}.

\section{A Geometric formulation of the SDC}
\label{sec:subspaces}

Let us recap our approach in general language. Consider some theory satisfying the SDC, i.e. containing towers of states decaying exponentially for every geodesic approaching an infinite distance limit of the moduli space $\mathcal{M}$. When adding a scalar potential lifting some of the directions, we will be left with a new moduli space $\overline{\mathcal{M}}$  whose geodesic trajectories might lift to non-geodesic trajectories in $\mathcal{M}$. To satisfy the SDC in the new IR theory, we need that the level of non-geodesicity of these trajectories is small enough to still allow for exponentially falling towers. Given the field metric of $\mathcal{M}$ and the towers of states, we can always identify the non-geodesic trajectories that are consistent with satisfying the SDC in the IR theory, where the limiting cases are dubbed critical paths. This was done in Section \ref{sec:hyp} for the case of products of hyperbolic planes, obtaining specific bounds for the critical paths. However, this procedure requires specific information about the geometry of the moduli space, so it needs to be worked out case by case. 

In this section, we are going to take a step back and reformulate the SDC in a language that will allow us to generalise the results of section \ref{sec:hyp} for general moduli spaces. This will be later translated into a Convex Hull condition  in analogy to the WGC in section \ref{sec:convex-hull}. The strategy is to provide a geometric description of the criteria for trajectories to fulfill or violate the SDC. We will keep the nomenclature introduced in section \ref{sec:hyp} to distinguish between the different types of trajectories, namely:
\begin{itemize}
\item Asymptotically geodesic paths: they approach a geodesic in the asymptotic limit, so the exponential rate is that of the geodesic.
\item Critical paths: non-geodesics that still marginally allow for the exponential decay of the tower, although the exponential rate differs from the geodesic one.
\item Swampy paths: they highly deviate from geodesics so the tower does no longer fall exponentially.
\end{itemize}

\smallskip

\subsection{Geometric formulation}
\label{sec:geometric-formulation}

The general formulation of the SDC establishes that, for any geodesic in moduli space in an infinite distance limit  there exists a tower of states with mass scale
\beqa
M\,=\, \exp \,(\, -\alpha\, \Delta\,)
\eeqa
with positive $\alpha>0$. Stronger versions of the conjecture further require $\alpha>\mathcal{O}
\left( 1\right)$, motivated by string theory setups. Let us now focus in the mildest version and simply require $\alpha>0$, leaving the study of the implications of $\alpha>\mathcal{O}
\left( 1\right)$ for the next subsection.

Consider a trajectory $\gamma$ approaching an infinite distance point in moduli space. We start by rewriting the exponential decay rate $\alpha$ of the tower mass scale as
\begin{equation} 
\label{exp-rate}
	\alpha(\Delta) = - \frac{d\log M}{d\Delta} \,=\, -T^{i}\, \partial_{i} \log M \, ,
\end{equation}
where $T$ is the normalized tangent vector of $\gamma$, and $M$ is implicitly evaluated along it.

This rewriting shows that the only information about $\gamma$ relevant for the SDC is the limiting tangent vector when approaching the infinite distance point. This agrees with our observation that the modulus of the proper acceleration \eqref{proper-acceleration} is not the right quantity to discriminate the behaviour of asymptotic trajectories. 

An important observation is that the set of allowed asymptotic tangent vectors $T$ near a point at infinity is in general restricted, in particular for asymptotically geodesic trajectories.\footnote{Notice that for this it is crucial that the point at infinity is singular. For a regular point, space is locally flat and thus any tangent vector corresponds to a geodesic passing through it.\label{foot:singular}} For instance, in the hyperbolic plane case, any curve approaching $s\to\infty$ with bounded $\phi$ must have $\dot{\phi}\to 0$; hence only the asymptotic tangent vector in the $s$ direction is allowed. On the other hand, relaxing the asymptotic geodesicity requirement allows to explore more general vectors $T$, as in the case of critical or swampy trajectories. To reflect this fact, we define the subspace $\mathbb{G}$ as that spanned by asymptotic tangent vectors of asymptotically geodesic trajectories.\footnote{Strictly speaking, we are interested in the subset containing all asymptotically geodesic vectors, which may not necessarily form a vector subspace. However, being it the case in all the examples at hand motivated by string theory, we will treat $\mathbb{G}$ as a well-defined vector subspace.}

Let us now consider the implications of the SDC (in its milder version) for such asymptotically geodesic trajectories. We note that what appears in \eqref{exp-rate} is the scalar product between the (limit) tangent vector and the gradient of $\log M$. 
This implies that a single tower of states along an asymptotically geodesic direction suffices to satisfy the SDC for any other direction, except for the orthogonal ones. Thus, the minimal requirement of the SDC is that there exist as many towers as orthogonal limit tangent vectors in $\mathbb{G}$. 

Actually, in general, there may exist other towers of states beyond the above minimal set. Hence it is convenient to consider a new subspace, denoted by $\mathbb{M}$, spanned by the gradient vectors of (log of) the scale $M$, for all existing towers of states. Note that in many string theory realizations the directions associated to such towers are ``dense''; for instance, in a KK compactification on $\IS^1\times \IS^1$ near the decompactification limit $R_1,R_2\to\infty$, there are towers of KK states with masses
\beqa
M^2\,=\, \Big(\, \frac{n_1}{R_1}\, \Big)^2\, +\, \Big(\, \frac{n_2}{R_2}\, \Big)^2\ .
\eeqa
Hence, for any rational direction of  $\gamma$ i.e. $R\equiv R_1/q_1=R_2/q_2$, there is a tower of states with mass $M\sim n/R$ by taking the states $n_1=nq_1$, $n_2=nq_2$.

In terms of these spaces, the mildest version of the SDC (with $\alpha>0$) can be expressed as: \\

\noindent\fbox{
\begin{minipage}{6in}
For any vector in $\mathbb{G}$ (i.e. any asymptotically geodesic tangent vector),
there must be at least one non-orthogonal vector in $\mathbb{M}$ (i.e. a suitable tower of states becoming massless). 
\end{minipage}}\\

\noindent Equivalently, the projection of $\mathbb{M}$ onto $\mathbb{G}$ should completely fill the latter:
\begin{equation} \label{SDC-subspaces}
	\mathcal{P}_{\mathbb{G}} \mathbb{M} = \mathbb{G} \, .
\end{equation}
Incidentally, this implies that their dimensions satisfy  ${\rm dim}\,\mathbb{M}\geq {\rm dim}\,\mathbb{G}$.

This geometric formulation of the SDC resembles a kind of Completeness Hypothesis where the towers of states play the role of the charge spectra in gauge field theories. Analogously, here the role of the charge space is played by the space of asymptotic tangent vectors of asymptotically geodesic trajectories. Stronger conditions - similar to the WGC - will appear when further requiring the towers to satisfy a lower bound for the exponential rate $\alpha\geq \alpha_{0}$, with $\alpha_{0}$ some order one constant. 
This mild formulation, though, already allows us to extract interesting conclusions. The first observation is that a single tower of states might not suffice to satisfy the SDC whenever the space $\mathbb{G}$ is spanned by more than one tangent vector while $\mathbb{M}$ remains one-dimensional. This can occur e.g. in higher dimensional moduli spaces in which the tower misses to depend on at least one of the saxions. 
A second observation is that, by satisfying the above criterium, we are actually fulfilling the SDC along a more general set of trajectories beyond geodesics. We will characterize this set of trajectories in the next subsection.

\subsection{Non-geodesic bounds}
\label{sec:non-geodesic-bounds}

We  can now turn to characterizing which trajectories could satisfy or violate the SDC. 
Indeed, the presence of a single tower makes the mild version of the SDC satisfied for all its non-orthogonal directions, and not only geodesics. Taking into account all possible towers, this defines a subset $\mathcal{T}_{SDC}$ composed by all the directions that satisfy the SDC. In this way, the SDC can be reformulated as imposing that this subset must contain all the asymptotically geodesic directions,
\begin{equation} \label{eq:SDC-from-M_QG}
	\mathbb{G} \subset \mathcal{T}_{SDC} \, .
\end{equation}
Clearly, swampy trajectories violating the SDC will therefore correspond to those not belonging to $ \mathcal{T}_{SDC} $.
Notice that, for this mild version, $\mathcal{T}_{SDC}$ is just the whole set of directions in moduli space minus the orthogonal complementary of the subspace $\mathbb{M}$. Hence, critical trajectories also belong to $\mathcal{T}_{SDC}$ in this mild formulation.

Let us now turn to the stronger version of the SDC, in which we impose a lower bound for the exponential rate $\alpha\geq\alpha_{0}$ with $\alpha_0$ some $\mathcal{O}(1)$ constant. It is reasonable to assume that $\alpha$ cannot take arbitrarily small values as otherwise it would violate the exponential behaviour required by the SDC. Moreover, all string theory examples studied so far have $\mathcal{O}(10^{-1}-10)$, and precise bounds have been given in the context of towers of BPS particles in Calabi-Yau compactifications \cite{Grimm:2018ohb,Gendler:2020dfp}. There, one finds that $\alpha\geq \frac1{\sqrt{2n}}$ for a $CY_n$, implying $\alpha\geq \frac1{\sqrt{6}}$ for a $CY_3$ Type II compactification to four dimensions. 
A lower bound has also been motivated by using the Transplanckian Censorship Conjecture \cite{Bedroya:2019snp,Andriot:2020lea,Bedroya:2020rmd} or by identifying infinite distance limits with RG flow endpoints of BPS strings in 4d $\mathcal{N}=1$ EFTs \cite{Lanza:2020qmt}. Here, we will not commit  to any of these specific values for $\alpha_0$ although it would be extremely interesting to get a better understanding of this.

\begin{figure}[htb]
\begin{center}
\includegraphics[scale=.3]{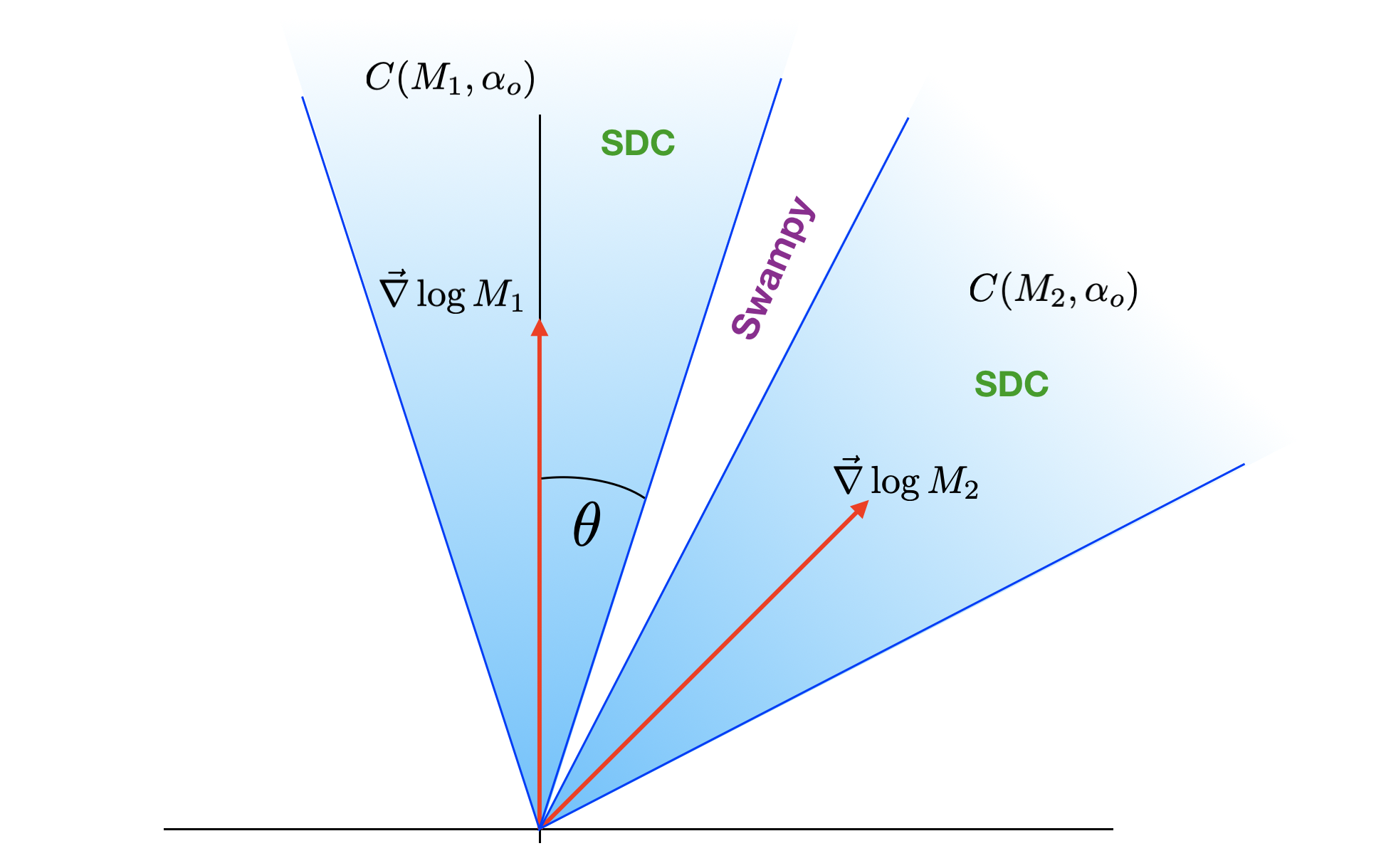}
\label{fig:cones1}
\caption{\small Pictorial representation of the subset of directions $\mathcal{T}_{SDC}$. There are two towers of states and their associated cones are represented. Every direction outside both of these cones does violate the SDC for $\alpha\geq\alpha_{0}$.}
\end{center}
\end{figure}

We can now extend the last formulation in \eqref{eq:SDC-from-M_QG} to include the lower bound on $\alpha$ by finding an appropriate definition of $\mathcal{T}_{SDC}$. To do this, let us recast the scalar product in \eqref{exp-rate} in terms of the angle $\theta$ between the (limit) tangent vector and the gradient of $\log M$, and get
\begin{equation} 
\label{exp-decay-angle}
	\alpha = - |\partial \log M| \cos \theta \, .
\end{equation}
Here we see that a single tower will make the SDC with $\alpha\geq\alpha_{0}$ satisfied for any direction such that
\begin{equation} \label{eq:max-angle}
	\cos \theta \leq - \frac{\alpha_{0}}{|\partial \log M|} \, .
\end{equation}
This is, each tower will then come with an associate cone of directions satisfying the SDC, $\mathcal{C}_{M}(\alpha_{0})$, and defined by \eqref{eq:max-angle}. Therefore, $\mathcal{T}_{SDC}$ will be formed by the union of the associate cones of all the towers of states (see figure \ref{fig:cones1}), this is,
\begin{equation}
\label{TSDC}
	\mathcal{T}_{SDC} = \bigcup \mathcal{C}_{M_i}(\alpha_{0}) \, .
\end{equation}
With this definition, the SDC with $\alpha\geq\alpha_{0}$ reduces again to \eqref{eq:SDC-from-M_QG}. Notice that not all critical trajectories will satisfy \eqref{eq:max-angle}, so only a subset of them will belong to $\mathcal{T}_{SDC} $, whose definition now depends on $\alpha_0$. We will translate this condition into a convex hull condition in Section \ref{sec:convex-hull}, which provides an equivalent but simplified and more elegant formulation of the above criterium. 

In general, determining $\mathcal{T}_{SDC}$ is not only associated to the non-geodesicity of the trajectory but requires full information about the tower of states. However, it becomes a purely geometric condition in the particular case that
\begin{equation} \label{strongSDC}
	\mathbb{M} = \mathbb{G} \, .
\end{equation}
In this situation, the realization of the SDC is such that the non-geodesicity of trajectories is directly related to the slow-down of the exponential falloff of state masses along them. It is in this case when the moduli space geometry completely determines the structure of the towers near infinity, as indeed occurred in earlier examples, and in most string theory examples. Hence, it is a natural framework to discriminate the asymptotically geodesic, critical and swampy trajectories, generalizing our discussion in Section \ref{sec:hyp}.

Our approach allows to go even further, and also obtain the modification factor $\nu$ in the exponential decay rate for the non-geodesic cases, as follows. 
The exponential rate in \eqref{exp-rate} can be written as
\begin{equation}
	\alpha = -(\mathcal{P}_{\mathbb{M}}T)^{i} \partial_{i} \log M=(\mathcal{P}_{\mathbb{G}}T)^{i} \partial_{i} \log M \, .
\end{equation}
where we have used \eqref{strongSDC} in the last step. Notice that $\mathcal{P}_{\mathbb{M}}T$ is nothing else than $\cos(\theta)$ defined in \eqref{exp-decay-angle}.
For a unit vector in $\mathbb{G}$, the result of this expression is the non-vanishing exponential decay rate required to fulfil the SDC along geodesics, i.e.
\begin{equation}
	\alpha = |\mathcal{P}_{\mathbb{G}}T| \, \alpha_{\text{geod}}.
\end{equation}
Hence, the factor $\nu$ in \eqref{eq:nu-factor} is given by
\begin{equation} \label{gen-criterion}
	\nu = |\mathcal{P}_{\mathbb{G}}T|^{-1} = (1 -
	|\mathcal{P}_{\mathbb{G}^{\bot}}T|^2)^{-1/2} .
\end{equation}

\medskip

It is straightforward to apply these concepts to recover the results for the hyperbolic space in Section \ref{sec:hyperbolic}. As can be readily checked from (\ref{hyper-tower}), the relevant subspaces are spanned by $\partial_s$,
\beqa
\mathbb{G}\, =\, \mathbb{M}\, =\, \langle \,\partial_s\, \rangle\, .
\eeqa
For trajectories $\phi=f(s)$ we have the tangent vector
\beqa
T = \frac{s}{n\sqrt{1+f^{\prime}(s)^2}} \left(\partial_{s} + f^{\prime}(s) \, \partial_{\phi}\right) \, .
\eeqa
Non-geodesic trajectories are those with a nontrivial $\partial_\phi$ component in their limit tangent vector. Finally, the criterion depending on the modulus of the proper acceleration (\ref{eq:RSDC-violation}) is recovered from \eqref{gen-criterion} by checking, in the limit $s\to\infty$, the relation
\begin{equation}
	\mathcal{P}_{\mathbb{G}^{\bot}}T = n |\Omega| .
\end{equation} 

One can similarly recover the results for products of hyperbolic spaces from these considerations, as the interested reader in encouraged to check. We instead move on to provide an even more intuitive formulation of these criteria in terms of a Convex Hull condition similar to that used for WGC.

\section{The Convex Hull SDC}
\label{sec:convex-hull}

In this Section we formulate the SDC in terms of a Convex Hull condition in the space of asymptotic trajectories. This will let us to easily recover our earlier results about different classes of asymptotic trajectories in a pictorial way which is more familiar in the Swampland program. Moreover, it will also allow us to generalize the story for any combination of towers of states and any asymptotic structure of the field space.

\subsection{General formulation}
\label{sec:formulation-sdc}

Consider a trajectory $\gamma$ approaching an infinite distance point in moduli space and $\vec{T}$ its normalised tangent vector. As in section \ref{sec:subspaces}, we denote by $\mathbb{G}$ the subspace spanned only by asymptotically geodesic vectors, i.e. that approach a geodesic trajectory at infinite distance.  We have seen that requiring the existence of an infinite tower of states becoming light along any of these asymptotically geodesic trajectories, actually allows for satisfying the SDC along a more general set of trajectories characterised by vectors in $\mathcal{T}_{SDC}\supset \mathbb{G}$.  This larger space  allows for a certain level of non-geodesicity, including  critical paths but excluding swampy trajectories, according to the nomenclature summarised at the beginning of section \ref{sec:subspaces}. If we require a stronger version of the SDC in which the exponential rate in \eqref{exp-rate} satisfies a lower bound $\alpha\geq \alpha_0$, only a subset of the critical paths will be included in $\mathcal{T}_{SDC}$. In other words, the SDC with $\alpha\geq\alpha_{0}$ is equivalent to requiring that, for any direction in $\mathbb{G}$, there must exist a tower of states such that the gradient vector of $\log M$ projected onto that direction is sufficiently large. Our goal now is to translate this statement into a convex hull condition.


The key observation is that there is a formal analogy with WGC quantities. The gradient  of $ M$ can be regarded as the scalar charge of the tower under the moduli. We can also think of $\mathbb{G}$ as the vector space of possible `charge' directions. Hence, the previous criterion can be rephrased as requiring that \emph{for every charge direction, there must exist a charged infinite tower of states satisfying $\alpha(\Delta)\geq\alpha_0$ asymptotically}, where $\alpha_0$ is a fixed contant (of order 1) which quantifies the criterion of fast enough decay to satisfy the SDC.

The SDC conditions can thus be formulated in analogy with the scalar version of the WGC \cite{Palti:2017elp}. In particular for a tower with (scalar dependent) mass scale $M$, we can define a scalar charge to mass ratio
\beqa
\vec{z}\,=\, - g^{-\frac 12} \, \vec{\nabla}\, \log M\, ,
\eeqa
where $g^{1/2}$ is a matrix whose square is the field metric (more precisely, introducing the $n$-vein $e_i^ae_j^b \delta_{ab}=g_{ij}$, we have $z^a=-e^a_i g^{ij} \partial_j \log M$). The inclusion of the metric absorbs a piece in (\ref{exp-rate}), such that scalar products become cartesian in the following.

The scalar WGC requires the existence of at least one state satisfying $|\vec{z}|\geq \mathcal{O}(1)$, such that the gravitational force acts weaker than the scalar force \cite{Palti:2017elp} (see \cite{Gonzalo:2020kke} for a different motivation of this proposal). Hence, the order one factor is typically fixed such that states saturating the scalar WGC should feel no force. Unlike with the usual WGC, the order one factor is not associated to extremality of black holes but, for convenience, we will keep the terminology \emph{extremal} to refer to those states saturating the bound. At first glance, it seems that the scalar WGC is different to the SDC, as for the latter what matters is not the modulus of the scalar charge to mass ratio but the projection over a trajectory. However, we will se that the SDC can actually be understood as a Convex hull Scalar WGC in which the extremal states are instead identified as those decaying exponentially with a minimum rate $\alpha_0$.

\begin{figure}[htb]
\begin{center}
\includegraphics[scale=.5]{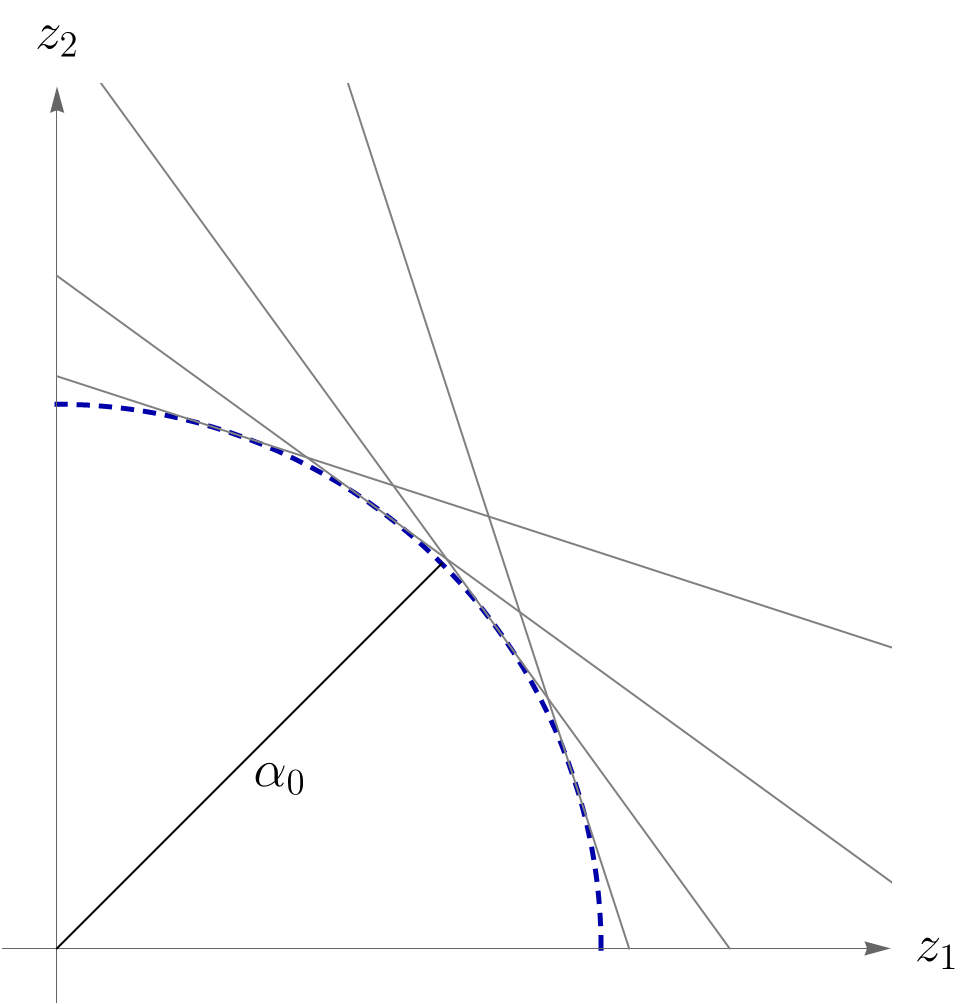}
\label{fig:extremal}
\caption{\small The extremal region as envelope of hyperplanes.}
\end{center}
\end{figure}

Consider a vector space  of dimension equal to the number of scalars under consideration, and a general unit vector $\vec{n}$ therein to parametrize the asymptotic behaviour of a general trajectory. This is related to the earlier vector $\vec{T}$  by $n^a=e_i^aT^i$, and is unit norm with respect to the Cartesian dot product. We define the \emph{extremal states} as those with a scalar charge to mass ratio vectors $\vec{z}$ satisfyng
\beqa
\vec{n}\cdot\vec{z}=\alpha_0
\label{the-dot-condition}
\eeqa
for some fixed $\alpha_0>0$ determining the lower bound for the SDC exponent. From (\ref{exp-rate}), we see that for fixed $\vec{n}$, this corresponds to the full set of towers with exponential rate $\alpha_0$ along the asymptotic trajectory defined by $\vec{n}$. It corresponds to a hyperplane orthogonal to $\vec{n}$, at a distance $\alpha_0$ from the origin. Scanning over all possible unit vectors,\footnote{Note that we allow for both positive and negative values of all components of $\vec{n}$. This is unphysical for the scalars becoming large in trajectories going off to points at infinity. However, we allow for this possibility at the formal level, to produce a simpler formulation of the convex hull condition, out of which the physical constraints follow from simple restriction of the allowed trajectories.
\label{the-image}} we define the {\em extremal region} as the enveloping hypersurface defined by the set of all such hyperplanes. It corresponds to a sphere or radius $\alpha_0$, see Figure \ref{fig:extremal}. 

By allowing the sphere in Figure \ref{fig:extremal} to take any radius, we recover the mildest version of the SDC in which $\alpha$ is an undetermined positive constant, $\alpha > 0$. 
However, it seems reasonable to consider a finite radius for the extremal ball which cannot be taken parametrically small, as that would spoil the exponential behaviour and violate the SDC. Determining how small $\alpha$ can get is one of the biggest open questions of the SDC and, as explained above, specific lower bounds have been proposed in the literature \cite{Grimm:2018ohb,Bedroya:2019snp,Andriot:2020lea,Gendler:2020dfp,Bedroya:2020rmd,Lanza:2020qmt}. 
One possibility motivated by the key role of the scalar charge to mass ratio above is that $\alpha_0$ can indeed be determined by using the scalar WGC or some sort of no-force requirement, as it was probably envisioned in \cite{Palti:2017elp}. This would be very interesting as it might be used to provide a bottom-up rationale for the SDC.

It is now straightforward to define the SDC in terms of a convex hull\footnote{To achieve a full convex hull, we formally use the method of images and also include the mirror vectors along the negative directions mentioned in footnote \ref{the-image}.} condition:\\

\noindent\shadowbox{
\begin{minipage}{6in}
\textbf{Convex Hull SDC:} In a theory with a set of towers corresponding to scalar vector to charge ratios $\vec{z}_i$, the requirement that the SDC is satisfied (with at least decay rate $\alpha_0$) by any trajectory is exactly the condition that the convex hull of the vectors $\vec{z}_i$ contains the above defined extremal region, namely the unit ball of radius $\alpha_0$. 
 \end{minipage}}\\
 
 Alternatively, it is possible that the the SDC convex hull condition is not satisfied, so the SDC does not hold (with decay rate $\alpha_0$) for all trajectories, but it still applies to some trajectories. In this situation we can put bounds on the trajectories not to become swampy, constraining the level of non-geodesicity allowed such that the SDC is satisfied. This situation naturally occurs when we start with a UV theory satisfying the SDC and then add a scalar potential lifting some directions, so we are left with a  IR moduli space whose geodesics might lift to non-geodesics from the UV perspective. In this case, we can use the convex hull SDC in the UV theory to constrain the allowed set of non-geodesic trajectories that would still allow us to comply with the SDC in the IR. The advantage of this formulation is that it also allows us to incorporate the possibility that new towers of states appear in the IR theory when adding the scalar potential.
  Hence, the Convex Hull SDC can be used to constrain either:
 \begin{itemize}
 \item the spectra of the theory, by requiring as many towers as needed to satisfy the convex hull condition,
 \item or the possible trajectories along which the SDC can be satisfied for a fixed set of towers and, therefore, the scalar potentials consistent with quantum gravity.
 \end{itemize}
 This latter option is not possible in the usual WGC, as the charge lattice is typically a fixed input of the theory.\footnote{Actually, the charge lattice can vary after higgsing a gauge group. If the higgsing is too large, this can lead to a violation of the WGC in the IR, even if it was originally satisfied in the UV. This is a known loophole of the WGC \cite{Saraswat:2016eaz}. From our perspective, by analogy with the SDC, the resolution is that the the amount of higgsing should be restricted in quantum gravity, so the WGC could also be used to constrain the allowed IR charge lattices.} However, it is very natural in the context of the SDC, as the allowed set of trajectories consistent with quantum gravity is still an open question, as it depends in turn on what scalar potentials can be realised in quantum gravity.

 
\subsection{Examples}
\label{sec:chsdc-examples}

In this Section we illustrate these ideas with examples, reproducing and generalizing the results in previous sections. 

\subsubsection{The hyperbolic plane complex scalar revisited\label{sec:ex1}}

Consider the case of the hyperbolic plane in section \ref{sec:hyperbolic}. Using the metric (\ref{hyperbolic-plane}) the charge to mass ratio vector for a tower with mass scale $M$ is
\beqa
\vec{z}\,=\, - \frac sn\, (\partial_\phi \log M, \partial_s \log M)\, .
\eeqa

Asymptotically geodesic trajectories have a tangent vector which approaches $\vec{n}=(0,1)$ asymptotically. Critical trajectories are parametrised by (\ref{single-path}) with constant $f' = \beta$, so the unit vector is
\beqa
\vec{n}\,=\, \frac{1}{\sqrt{1+\beta^2}}\, (\beta,1)\, .
\label{vector-n-one}
\eeqa
For these trajectories the vector $\vec{n}$ is constant so that (\ref{the-dot-condition}) corresponds to the equation of a straight line in the plane $(z_1, z_2)$. Different trajectories with different values of $f'$ will give rise to different straight lines, e.g. horizontal lines correspond to a purely saxionic trajectory, and bigger $\beta$ leads to bigger slopes.

\begin{figure}[htb]
\begin{center}
\includegraphics[scale=.5]{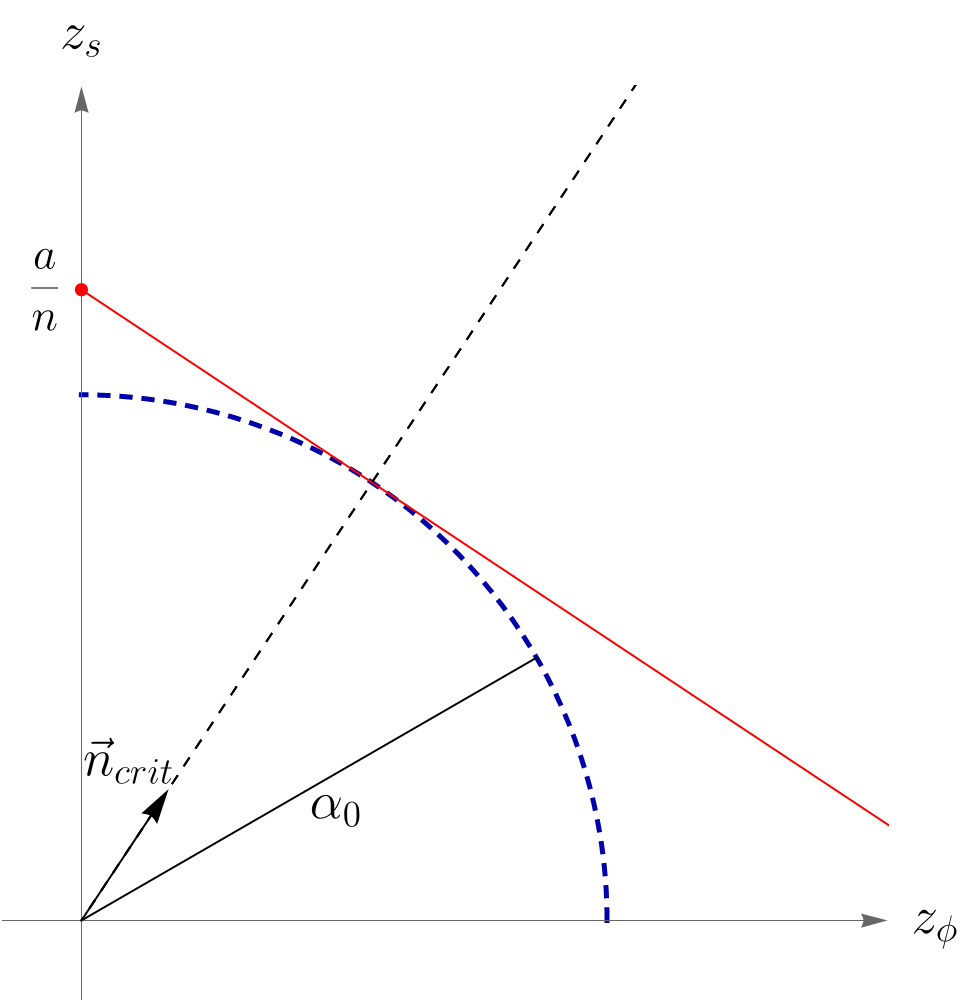}
\caption{\small The bound on almost saxionic trajectories.}
\label{fig:almost-saxionic}
\end{center}
\end{figure}

For the particular case of a single tower $M\sim s^{-a}$, c.f. (\ref{hyper-tower}), we have a single point (and its image) at $\vec{z}=(0,\pm a/n)$. Clearly, the convex hull of these two points does not contain the ball of radius $\alpha_0$, hence it does not satisfy the SDC for any trajectory. The SDC is satisfied only in the purely saxionic (geodesic) direction if $a/n>\alpha_0$, or trajectories close enough to it. We can then use the convex hull condition to put a bound of how much a trajectory can deviate from the geodesic saxionic trajectory. For this porpuse, we just need to compute the  angle $\cos\theta=1/\sqrt{1+\beta_{\rm max}^2}$ at which a tangent trajectory to the ball passes by the point $\vec{z}=(0,\pm a/n)$ (see Figure \ref{fig:almost-saxionic}). This occurs for a trajectory with
\beq
\beta_{\rm max}=(\cos\theta)^{-2}-1=\left(\frac{a}{n\alpha_0}\right)^{2}-1
\eeq
Hence, critical trajectories with $\beta\leq \beta_{\rm max}$ will satisfy the SDC with a exponential rate given by 
\beq
\alpha_{\rm crit.}=\frac{a}{n\sqrt{1+\beta^2}}\, ,
\label{crit-alf}
\eeq
recovering the result (\ref{critical-values}) in section \ref{sec:hyperbolic}.

\begin{figure}[htb]
\begin{center}
\includegraphics[scale=.45]{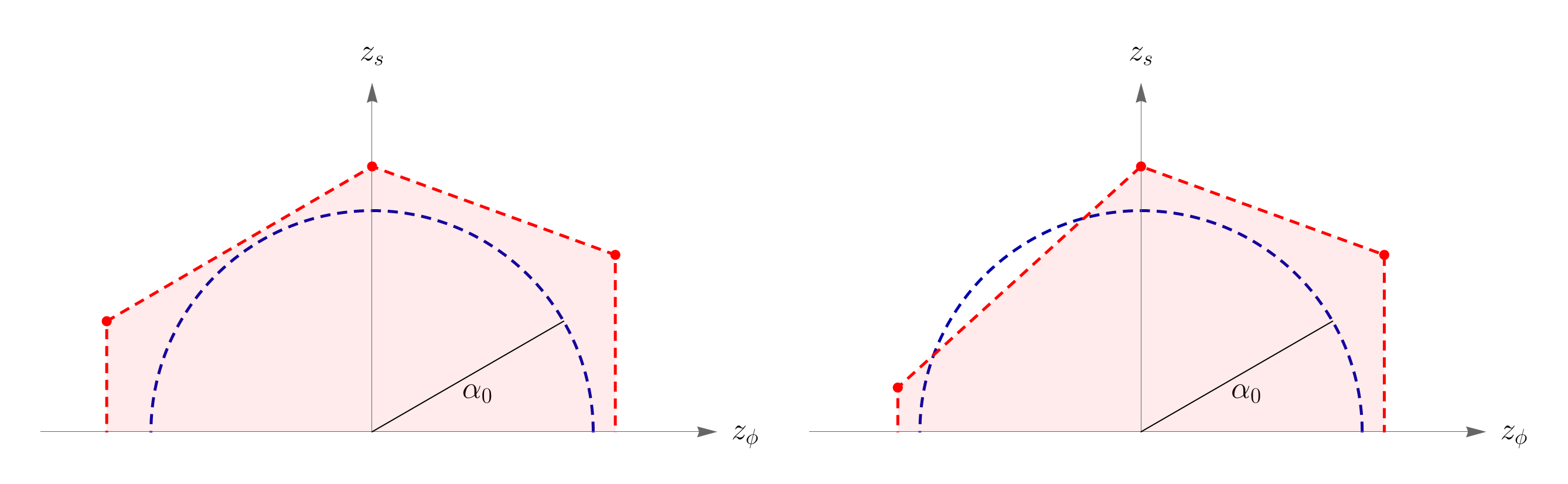}
\caption{\small The convex hull satisfied or not.}
\label{fig:yes-not}
\end{center}
\end{figure}

\begin{figure}[htb]
\begin{center}
\includegraphics[scale=.3]{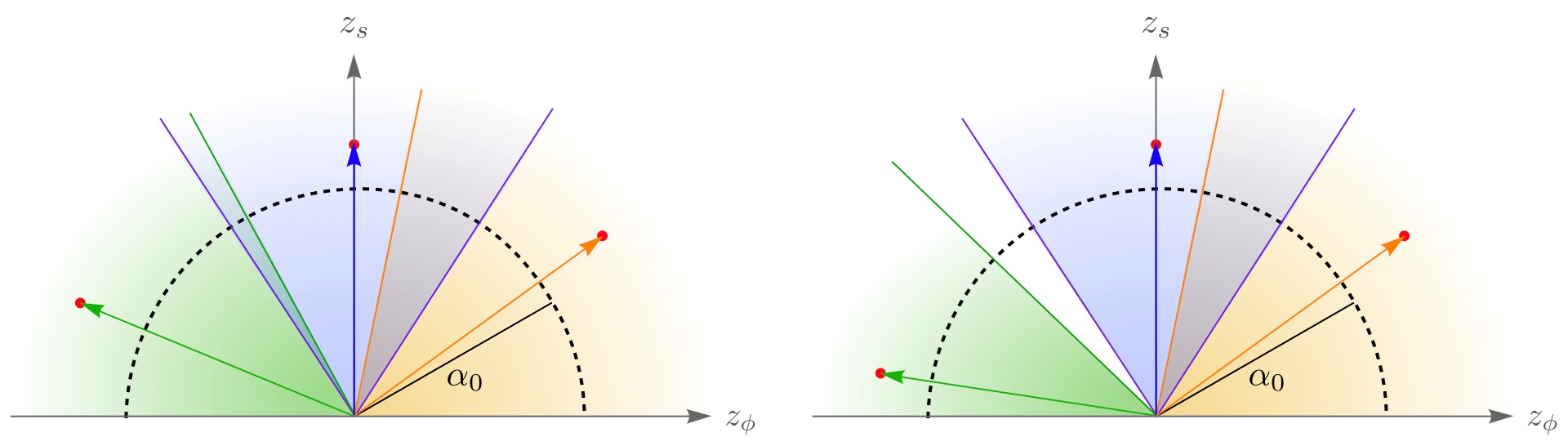}
\caption{\small Same setups as those shown in Figure \ref{fig:yes-not} from the perspective of the subset $\mathcal{T}_{SDC}$.}
\label{fig:cones2}
\end{center}
\end{figure}

Alternatively, if one is interested in enforcing the SDC for any trajectory, we have to introduce more towers, such that the convex hull of their $\vec{z}$'s encloses the extremal region. In Figure \ref{fig:yes-not} we depict situations with different towers and fulfilling, or not, the SDC for any trajectory. It is instructive to compare with the criterion in section \ref{sec:geometric-formulation} in terms of the cones comprising $\mathcal{T}_{SDC}$, see Figure \ref{fig:cones2}.

\subsubsection{Two saxions\label{sec:ex2}}

Let us consider now a theory with two saxion-like real scalars, namely with a metric
\begin{equation}
    d\Delta^{2}=\frac{n_1^{2}}{s_1^{2}}\,ds_1^{\,2}\,+ \,\frac{n_2^{2}}{s_2^{2}}\,ds_2^{\,2}\, .
    \label{two-saxion-metric}
\end{equation}
This can be considered as template for the situation with two complex scalars, with hyperbolic space metric (\ref{two-hype-metric}), if we restrict to trajectories not involving the corresponding axions.

The scalar charge to mass ratio for a general tower with mass scale $M(s_1,s_2)$ is
\beqa
\vec{z}\,=\, - \big(\,\frac{s_1}{n_1}\, \partial_{s_1}\, \log M\, , \frac{s_2}{n_2}\, \partial_{s_2}\, \log M\, \big)\, .
\eeqa
A typical situation is to have two towers, each ensuring the SDC along its corresponding saxionic direction
\beqa
M_1\,\sim\, s_1^{-a_1}\quad,\quad M_2\,\sim\, s_2^{-a_2}\, .
\eeqa
%
\begin{figure}[htb]
\begin{center}
\includegraphics[scale=.4]{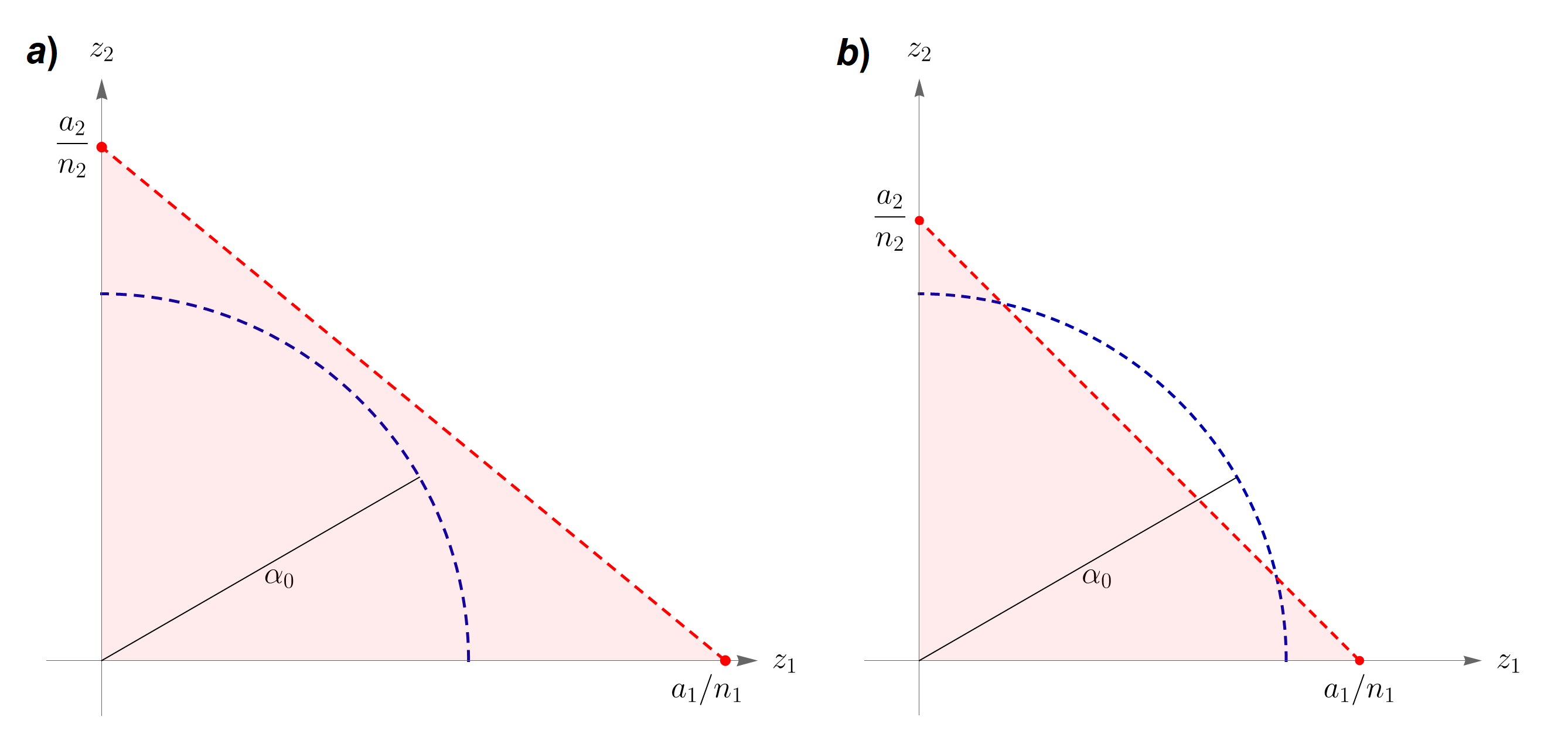}
\caption{\small The convex hull satisfied or not for two saxions, depending on the specific values of $a_i,n_i$. For simplicity, we only show the positive quadrant.}
\label{fig:saxion-triangles}
\end{center}
\end{figure}
%
This corresponds to the values $\vec{z}_1=(a_1/n_1,0)$ and $\vec{z}_2=(0,a_2/n_2)$ respectively. In Figure \ref{fig:saxion-triangles} we depict some examples of the corresponding convex hull conditions. Note that even if the SDC is satisfied along each saxion direction individually, it may fail along some other mixed trajectories, see Figure \ref{fig:saxion-triangles}b. This is reminiscent of similar behaviours in the WGC, see e.g. \cite{Montero:2015ofa}. The condition that the SDC is satisfied (with decay rate $\alpha_0$) for any trajectory is straightforward to get from the geometric figure:
\beqa
\frac{a_1 a_2}{n_1n_2}\,\bigg[\,\bigg( \frac{a_1}{n_1}\bigg)^2+\bigg( \frac{a_2}{n_2}\bigg)^2\bigg]^{-\frac 12}>\, \alpha_0\, .
\eeqa

It is interesting to compare this with the case in which the states are, or are not, mutually BPS. For instance, if we consider that the two towers of states are mutually BPS and can form threshold bound states, we expect there are towers with mass scales
\beqa
M\,=\, q_1\,M_1\,+\,q_2\,M_2 \, .
\label{sum-rule}
\eeqa
For these, the scalar charge to mass ratio is given by
\beqa
\vec{z}_{q_1,q_2}\, =\, \,\Big(\, \frac{a_1}{n_1}\, \frac{q_1}M\, s_1^{-a_1},\frac{a_2}{n_2}\, \frac{q_2}M\, s_2^{-a_2} 
\,\Big) \, .
\eeqa
Denoting its two components $\vec{z}=(z_1,z_2)$, they all lie in the hyperplane
\beqa
\frac{n_1}{a_1}\, z_1\, +\, \frac{n_2}{a_2}\, z_2\, =\, 1 \, ,
\eeqa
which is the line joining the two towers, namely the red dashed line in Figure \ref{fig:saxion-triangles}.
Hence, mutually BPS states do not need to comply with our definition of extremal in the context of the SDC convex hull. This can be important in the case in which the towers correspond to excitation modes of mutually BPS strings, as e.g. in \cite{Lanza:2020qmt}. It also contrasts with the usual WGC, where only BPS states are expected to have a \emph{gauge} charge to mass ratio equal to an extremal black hole.
It would be interesting to clarify the interplay of the two notions of extremality to possibly improve on the SDC convex hull formulation.

On the other hand, if we consider towers of states which are not mutually BPS, they might form an ellipse in the scalar charge to mass ratio plane, satisfying the SDC convex hull condition more easily. This can play an important role when checking the SDC in the context of BPS towers of particles in CY compactifications and patching the results from different growth sectors.\footnote{Each growth sector corresponds to an specific ordering of the saxions, regarding which one grows faster when approaching the infinite distance loci. Typically, the tests of the SDC in this context focus on each growth sector independently \cite{Grimm:2018cpv,Gendler:2020dfp}.}

\subsubsection{Decoupled Saxion-Axion\label{sec:ex3}}

In this section we consider trajectories involving a saxionic scalar $s$, and an axionic scalar $\psi$ corresponding to a {\em different} saxion $u$, namely the metric reads
\beqa
d\Delta^2\,=\,\frac{n^2}{s^2}\, ds^2\,+\,\frac{m^2}{u^2}\, d\psi^2\, .
\eeqa
Clearly the prefactor of the second term can be removed by redefining $\psi$, but we prefer to keep it. This allows an easier interpretation of the results as a subsector in a model with two hyperbolic plane complex scalars, c.f.(\ref{two-hype-metric}).

The scalar charge to mass ratio has the form
\beqa
\vec{z}\,=\, - \big(\, \frac um\,\partial_\psi\log M, \frac sn\,\partial_s\log M\, \big) \, .
\eeqa

Asymptotically, the trajectory can be parametrized with $s$, so it reads $\psi=f(s)$. The corresponding unit vector is
\beqa
\vec{n}\,=\, \frac{1}{\sqrt{\frac{m^2}{u^2}f'{}^2+\frac{n^2}{s^2}}}\,\Big(\,\frac mu f' , \frac ns
\,\Big)\,=\,
\frac{1}{\sqrt{\frac{m^2}{n^2}\frac{s^2}{u^2}f'{}^2+1}}\, \Big(\,\frac mn \frac su f' , 1 \,
\,\Big) \, ,
\eeqa
where the last equality is just a convenient rewriting.
As derived in section \ref{subsec:two-hype}, critical trajectories in this subsector (with $u$ constant) obey $sf'\to \gamma={\rm const.}$, yielding $f(s)\,\to\, \gamma\,\log s $. For these trajectories, the unit vector reads
\beqa
\vec{n}=\frac{1}{\sqrt{1+\beta^2}}(\beta,1) \, ,
\label{vector-n-two}
\eeqa
where we have defined $\beta=\frac m{nu} \gamma$. If $f(s)$ grows faster, we recover $\vec{n}=(1,0)$ (swampy paths), while if it grows slower we obtain $\vec{n}=(0,1)$ (asymptotically geodesic paths).
As in the previous examples, \eqref{vector-n-two} scans over different directions in the 2d plane of $\vec{z}$, covering all possible critical trajectories.

It is straightforward to consider different possible towers and analyze whether the Convex Hull SDC is satisfied, or else, which bounds it sets on the parameters of the model and the allowed trajectories. For instance, since (\ref{vector-n-two}) is formally like  (\ref{vector-n-one}), if we consider a single tower with scaling  $M\sim s^{-a}$, we obtain a critical value of the decay rate along the trajectory (\ref{crit-alf}).
In other words, 
\beqa
\gamma\, =\, \frac{nu}m \,\sqrt{\frac{a^2}{n^2\alpha_{\rm crit}^2}-1} \, .
\label{crit-gam}
\eeqa
from which we can extract the value of $\alpha_{\rm crit}$ in terms of $\gamma$. Only along trajectories with $\gamma\leq \gamma(\alpha_{\rm crit}=\alpha_0)$ the SDC is satisfied.
This defines the maximal amount of excitation the axion $\psi$ can have not to spoil a given exponential decay rate $\alpha_{\rm crit}$ along the trajectory. 

Interpreting the result as applied to subsector of a two complex scalar model, a trajectory deviating from a geodesic single saxionic direction by exciting the axion of the second complex scalar preserves the SDC if the axion grows with at most the log dependence $\psi=f(s)\,\to\, \gamma\,\log s $ and $\gamma$ above. We leave to the interested reader the discussion of further possibilities of tower distributions and the corresponding bounds.

Combining the results of sections \ref{sec:ex1}, \ref{sec:ex2} and \ref{sec:ex3}, we complete the analysis of a two complex dimensional moduli space given by a product of hyperbolic planes.  This can be trivially generalised to products of more than two hyperbolic planes. As explained in section \ref{sec:asymptotic-flux}, they are good templates of the asymptotic geometry  realised at the infinite distance limits of Calabi-Yau compactifications.

\section{Constraints on the potential and asymptotic flux compactifications}
\label{sec:asymptotic-flux}

Throughout this paper, we have argued that consistency of the SDC at any energy scale put constraints on the set of nearly-flat field trajectories allowed by quantum gravity. This is because the moduli space of a theory, and consequently  the identification of geodesic paths, varies when going to the IR and integrating out heavy scalar degrees of freedom. But by placing bounds on the trajectories we are actually constraining the scalar potentials consistent with quantum gravity! In this section, we give some first steps translating our bounds to the potential and comparing with previous literature on the asymptotic behaviour of scalar potentials in string theory.

A natural setup in which to apply our above strategy is string theory flux compactifications. These are most often described by starting with a flux-less compactification, with a moduli space on which a potential is subsequently introduced by means of a flux superpotential. The resulting theory may maintain a moduli space of smaller dimension, if moduli stabilization is only partial, or the resulting potential may admit valleys which can be discussed as pseudomoduli. From our vantange point we are thus led to propose that the most general flux compactification must necessarily lead to potentials such that the resulting (pseudo)moduli space still satisfies the SDC. In particular, this implies that geodesics in this (pseudo)moduli space must belong to $\mathcal{T}_{SDC}$ defined in \eqref{TSDC}, and it should be impossible to get a valley along a highly turning trajectory which is not in  $\mathcal{T}_{SDC}$. We will see below that that in a fairly general class of models, the flux potentials precisely yield nearly-flat trajectories which are critical according to the definition at the beginning of section \ref{sec:subspaces}. In other words, the valleys of the potential have the maximum level of non-geodesicity (from the perspective of the original UV moduli space) that it is allowed to satisfy the SDC in the IR.

The asymptotic behaviour of the potential have been considered in quite some detail in CY flux compactifications in \cite{Grimm:2019ixq}. 
The setup is compactifications of M-theory on Calabi-Yau fourfolds with $G_4$ fluxes \cite{Becker:1996gj,Dasgupta:1999ss}, for which the mathematical machinery of asymptotic Hodge theory allows to study the asymptotic form of 
the flux potential near any infinite distance limit in complex structure moduli space. By taking the F-theory limit, one recovers a 4d $\mathcal{N}=1$ theory with a flux-induced scalar potential. This allows us to study, not only the more familiar infinite distance limits in pertrubative Type IIB/A, but also other types of limits for finite $g_s$. 
In the following, we summarize the results of  \cite{Grimm:2019ixq} that are relevant to our discussion, in order to reinterpret them from the new perspective advocated in this paper.

All infinite distance limits in complex structure moduli space of Calabi-Yau 
can be described as the loci of $\hat{n}$ intersecting complex divisors. In an 
appropriate parametrization, these are described by
\begin{equation}
	t^j = \phi^j + i s^j \to i\infty \, , \qquad j=0,\ldots,\hat{n}\, ,
\end{equation}
while all the other coordinates remain finite.
Taking $\phi^j$ and $s^j$ to be the axion and saxion of complex scalars, the above limits correspont to sending to infinity some of the saxion vevs. Using the Nilpotent Orbit Theorem \cite{schmid_variation_1973}, one can show (see e.g. \cite{Grimm:2018ohb,Grimm:2018cpv}) that the K\"ahler potential takes the following form in the asymptotic limit,
\beq
K=-\log (p_d(s^j)+\mathcal{O}(e^{2\pi i t^j}))
\eeq
where $p_d(s^j)$ is a polynomial of degree $d$ on the saxions, and $d$ characterizes the type of singular limit.\footnote{Although this also holds for singular loci at finite distance in moduli space, we restrict the discussion to infinite distance regimes, so $d\neq 0$.} More concretely, $d$ is associated to the properties of a monodromy transformation encoding the action of the axionic discrete shift symmetry in the limit.
For single moduli limits, i.e. $j=1$, the field metric exhibits the hyperbolic behaviour studied in Section \ref{sec:hyperbolic}:
\beq
d\Delta^{2}=  \frac{n^2}{s^{2}} 
	\left[ (ds)^{2}+(d\phi)^{2} \right]+d\Delta_{finite}^{2}
	\eeq
	with $n=d/4$ and $\Delta_{\rm finite}$ only depending on the moduli that are not taken to the  asymptotic limit.
The same behaviour occurs if we restrict to paths in some \emph{growth sector} in multi-moduli limits. This amounts to approaching the infinite distance limit in such a way that some axion vevs are much bigger than others. Namely, choosing as suitable ordering, we have $s^1 \gg s^2$, $s^2 \gg s^3$, ... and so on. In this so-called \emph{strict asymptotic regime} we can neglect  polynomial terms of the form $s^{j}/s^{j+1}$. The leading term of the K\"ahler potential can then be factorized,\footnote{For each saxion $s^j$, it is possible to define some integer $d_i$ characterizing the singularity. If all $d_i\neq 0$, then one simply has $n_i=d_i/4$. However, the factorization of $K$ breaks down when some $d_i=0$ and a more detailed analysis is required. We refer the reader interested in the details of these degenerate cases to \cite{Bastian:2020egp,LMMV}.} yielding
%
%
\begin{equation} \label{asymptotic-metric}
	d\Delta^{2}= \sum_{i} \frac{n_{i}^{2}}{(s^i)^{2}} 
	\left[ (ds^i)^{2}+(d\phi^i)^{2} \right] + \dots\, .
\end{equation}
to leading order in the asymptotic limit.
Namely, each complex modulus whose saxion is taken to the asymptotic limit parametrizes a hyperbolic plane, c.f. Section
\ref{subsec:two-hype}. 	Hereby our motivation to use hyperbolic metrics as toy models to illustrate our proposal and results in this paper.

Interestingly, not only the field metic, but also the flux-induced scalar potential is highly constrained in the infinite distance limits. In particular, it is possible to build an adapted basis of 4-cycles for each growth sector such that the corresponding cohomology group is divided in orthogonal subspaces under the Hodge norm in the strict asymptotic regime \cite{Grimm:2019ixq}. This induces a split of the $G_4$ flux in different components $G_4^\ell$, such that the scalar potential behaves as
$V\simeq \sum_\ell ||G_4^\ell||^2$.
Here $||G_4^\ell||^2$ denotes the Hodge norm of each flux component, whose moduli dependence can be completely determined using the discrete data characterizing the singular limit. Amusingly, the moduli dependence is such that the potential behaves as an homogeneous function\footnote{There were a few exceptions in \cite{Grimm:2019ixq} in which the potential was not homogeneous to leading order (see also \cite{LG}). However, they are not relevant for our analysis as they do not allow for parametrically large axionic field variations in a controlled regime with $s\gg 1$.} to leading order in the large field limit,
\beq
\label{V}
V(\lambda s^j,\lambda \phi^j)\simeq \lambda^mV(s^j,\phi^j)
\eeq

This was exploited in \cite{Grimm:2019ixq} to consider the question of the backreaction on the saxions due to the motion of the axion away from its minimum e.g. along an inflationary valley in an axion monodromy scenario \cite{Marchesano:2014mla,McAllister:2014mpa} (see \cite{Silverstein:2008sg,McAllister:2008hb,Kaloper:2008fb} for early axion monodromy models unrelated to flux compactifications).
 
As explained in \cite{Grimm:2019ixq}, the resulting backreaction for potentials satisfying \eqref{V} is of the form
\begin{equation}
 \label{linear-backreaction}
	 s \sim \beta \phi \, 
\end{equation}
with $\beta$ a flux-independent parameter. The above relation holds for each individual hyperbolic plane independently; namely, a trajectory in which the axions are excited away from their minima necessarily requires the saxions to have a backreaction linear in the corresponding axions. This implies that the valleys of the potential at the asymptotic regimes occur along \eqref{linear-backreaction}, so that e.g. highly turning axionic trajectories are not realised. Therefore, a tower of states decaying exponentially in the saxionic field, will also decay exponentially in terms of the axion, eventually signaling the EFT breakdown for large field variations.

This linear backreaction, and their correlation to the SDC, had previously been noted in certain models of Type II flux compactifications \cite{Baume:2016psm}, see also \cite{Valenzuela:2016yny,Blumenhagen:2017cxt,Buratti:2020kda}. It is also highly correlated to the difficulties for obatining mass hierarchies in these flux compactifications, as studied in \cite{Hebecker:2014kva,Blumenhagen:2014nba}. The analysis in  \cite{Grimm:2019ixq} supports that this behaviour is universal in flux Calabi-Yau compactifications, as it is tied to asymptotic properties of the moduli space inherited from Hodge Theory. But is it a general feature of potentials consistent with quantum gravity?

We are now ready to reinterpret \eqref{linear-backreaction} from a new perspective and provide an answer to the above question in view of the results of our paper. Noticing that the asymptotic moduli space metric is that of a hyperbolic plane, the linear result  (\ref{linear-backreaction}) corresponds to the critical case of non-geodesic trajectories in Section
\ref{sec:hyperbolic}. Namely, it corresponds to traveling along a trajectory which is as non-geodesic as possible in a way compatible with the distance conjecture. It is very exciting that string theory flux compactification thus saturate the non-geodesicity bound of the hyperbolic plane. It also implies that the flux potentials are consistent with the SDC being satisfied at any energy scale, providing evidence for our proposal.

Clearly, other asymptotic metrics could lead to different parametrizations of the critical paths. But the conclusion of our work is equivalent: the potential should be such that it only generates (pseudo)moduli spaces that ensure consistency of the SDC along the RG flow. This has interesting implications for single field inflation, including axion monodromy models. It would be interesting to turn the question around, and determine from a bottom-up perspective what is the more general form of the potential that generates nearly-flat trajectories corresponding to critical paths. In other words, such that the lightest field is associated to a field direction that coincides with a critical trajectory. One could then try to compare these general bounds on the potentials coming from the Convex Hull SDC with other swampland conjectures constraining the asymptotic form of the potential as the de Sitter conjecture \cite{Obied:2018sgi}.

Before closing this section, we would like to point out that a counterexample to the linear backreaction above was presented in \cite{Buratti:2018xjt} by considering field variations with a spatial dependence. There,  it was shown that the stabilization of the breathing mode is such that the resulting light mode avoids the KK tower to fall exponentially when approaching infinite distance. However, this does not contradict  our proposal, since this dangerous direction is not a geodesic from the perspective of the low energy pseudomoduli space. In other words, it belonged to the subspace $\mathbb{G}^{\bot}$ of the low energy pseudomoduli space, and thus the SDC was still satisfied in the IR. 

\section{Conclusions}
\label{sec:conclu}

In this paper we have discussed the interpretation of the Swampland Distance Conjecture in effective theories with scalar potentials leading to valleys of light fields. We have argued that the SDC is meaningful as a swampland constraint only if it applies at any scale, and that this poses non-trivial constraints of the potentials. We have approached the problem of characterizing these contraints by first studying the structure of non-geodesic trajectories near points at infinity in moduli spaces, and characterizing the constraints implied by the SDC. The analysis is carried out in hyperbolic spaces or products thereof, which provide a good template of general CY moduli spaces near infinite distance loci. We have shown that the critical behaviour of maximal non-geodesicity compatible with the SDC corresponds to axion variations with a linear backreaction on their corresponding saxions. We have argued that this agrees with the structure of flux compactifications near infinite distance loci. This suggests that string theory flux potentials are the most generic ones compatible with the SDC.

We have also reformulated the SDC in terms of a Convex Hull condition, in which scalar charge to mass ratio of SDC towers determine the exponential falloff $\alpha$ along asymptotic trajectories. The SDC is satisfied with an exponential rate lower bounded by $\alpha_0$ if the convex hull of the scalar charge to mass ratio of the towers includes the ball of radius $\alpha_0$. This allowed a very intuitive pictorial rederivation of the above mentioned results. For a given set of towers, it can be used to determine the set of trajectories consistent with the SDC, recovering the critical behaviour of maximum non-geodesicty above. Conversely, it can be used to argue for the existence of more than one tower in higher dimensional spaces.

\medskip

Our work opens several interesting avenues for future research. The existence of a Convex Hull SDC and the scalar charge to mass ratio are tantalizingly reminiscent of the WGC. It would be interesting to strengthen this connection in explicit string theory examples along the lines of \cite{Lee:2018spm,Gendler:2020dfp}. Moreover, we have seen that it is only upon requiring consistency of the SDC at any energy scale with a minimal exponential rate $\alpha_0$, that the SDC becomes formally equivalent to a convex hull Scalar WGC, implying bounds on non-geodesic trajectories. 
However, it is not clear whether the definition of the \emph{extremal} region should coincide for the SDC and the scalar WGC. Saturating the scalar WGC is associated to a no-force condition \cite{Palti:2017elp}, while the lower bound $\alpha_0$ for the SDC have been motivated based on the TCC \cite{Bedroya:2019snp,Andriot:2020lea,Bedroya:2020rmd} and string theory examples \cite{Gendler:2020dfp}. It would be interesting to understand better this lower bound for the SDC exponential rate and whether it can always been identified with saturating the scalar WGC or has a different origin. This might eventually provide a bottom-up rationale for the SDC.

We would also like to note that our results are in nice harmony with \cite{Lanza:2020qmt} in the context of BPS strings in $\mathcal{N}=1$ 4d EFTs. The gauge charge to mass ratio of these strings is equal to their scalar charge to mass ratio, and the string RG flow ensures that its tensions decreases exponentially with the distance. Hence, in this setup, the extremality factor for the WGC coincides with the scalar WGC and with the exponential decay rate. It is not surprising then that all these conditions look formally equivalent. 

It would also be interesting to use our ideas to produce constraints on general potentials in quantum gravity, and to compare them with existing swampland constraints, like the de Sitter conjectures \cite{Obied:2018sgi,Garg:2018reu,Ooguri:2018wrx} or the TCC \cite{Bedroya:2019snp}.
Another interesting source of potentials in string theory are non-perturbative D-brane instanton effects. These are relevant for attempts to achieve full moduli stabilization in type IIB models \cite{Kachru:2003aw}, and there are recent tools to recast them in terms of fluxes in a backreacted geometry \cite{Garcia-Valdecasas:2016voz} (see also \cite{Tenreiro:2017fon,Franco:2018vqd}). It would be interesting to verify the interplay of the resulting instanton flux contributions with the SDC. 
Finally, the derivation of the linear backreaction from the potential in asymptotic flux compactifications in \cite{Grimm:2019ixq} exploits homogeneity of the potential on the complexified moduli. This is also encountered in the action of certain $\IZ_k$ symmetries in \cite{Buratti:2020kda}. 
It would be interesting to explore the role of the discrete symmetries present in compactifications with large fluxes on the structure of their potential and hence on their mechanism to enforce the SDC.
We hope to come back to these questions in the near future.

\vspace{2cm}
%
\section*{Acknowledgments}
We are pleased to thank F. Baume, G. Buratti, L. Ib\'anez, F. Marchesano, M. Montero and M. Wiesner for useful discussions.
This work is supported by the Spanish Research Agency (Agencia Estatal de Investigaci\'on) through the grant IFT Centro de Excelencia Severo Ochoa SEV-2016- 0597, and by the grant PGC2018-095976-B-C21 from MCIU/AEI/FEDER, UE. The work by J.C. is supported by the FPU grant no. FPU17/04181 from
Spanish Ministry of Education. The research of I.V. was supported by a grant from the Simons Foundation (602883, CV).

\newpage

\bibliographystyle{JHEP}
\bibliography{mybib}

\end{document}